# Twofold Symmetry Observed in $Bi_2Te_3$/FeTe Interfacial Superconductor

Xinru Han[1, ‡], Hailang Qin[1, ‡], Tianluo Pan[1, 3, ‡], Bin Guo[1], Kaige Shi[1], Zijin Huang[1], Jie Jiang[1], Hangyu Yin[1], Hongtao He[1,2], Fei Ye[1,3], Wei-Qiang Chen[1,3], Jia-Wei Mei[1,3, *], Gan WANG[1,3, *]

[1]Department of Physics, Southern University of Science and Technology, Shenzhen 518055, China

[2]Shenzhen Key Laboratory for Advanced Quantum Functional Materials and Devices, Southern University of Science and Technology, Shenzhen 518055, China

[3]Guangdong Provincial Key Laboratory of Quantum Science and Engineering, Shenzhen Institute for Quantum Science and Engineering, Southern University of Science and Technology, Shenzhen 518055, China

**Abstract:**

**Superconducting pairing symmetry are crucial in understanding the microscopic superconducting mechanism of a superconductor. Here we report the observation of a twofold superconducting gap symmetry in an interfacial superconductor $Bi_2Te_3$/FeTe, by employing quasiparticle interference (QPI) technique in scanning tunneling microscopy and macroscopic magnetoresistance measurements. The QPI patterns corresponding to energies inside and outside the gap reveal a clear anisotropic superconducting gap. Furthermore, both the in-plane angle-dependent magnetoresistance and in-plane upper critical field**

**exhibit a clear twofold symmetry. This twofold symmetry align with the Te-Te direction in FeTe, which weakens the possible generation by bi-collinear antiferromagnetism order. Our finding provides key information in further understanding of the topological properties in Bi$_2$Te$_3$/FeTe superconducting system and propels further theoretical interests in the paring mechanism in the system.**

## Introduction:

Interface superconductors are of great interest and have their own unique territory, compared to other bulk superconductors or nano-island superconductors, in that the superconductivity can be naturally confined to a two-dimensional or quasi-two-dimensional space. There are a few interface superconductors found to date, such as PbTe/SnTe, LaAlO$_3$/SrTiO$_3$, La$_{1.55}$Sr$_{0.45}$CuO$_4$/La$_2$CuO$_4$, KaTiO$_3$/EuO, one unit-cell FeSe/SrTiO$_3$, Sb$_2$Te$_3$/FeTe and Bi$_2$Te$_3$/FeTe [1-7]. Among them, the superconductivity in the heterostructure of Bi$_2$Te$_3$/FeTe is particularly interesting owing to the fact that neither of the constituent material is superconducting; in addition, this system have the possibility to achieve 2D topological superconductor with spin-triplet pairing symmetry, due to the presence of both superconductivity and topological surface states at this interface based on the Fu-Kane model [8]. In similar systems, such as the Bi$_2$Te$_3$/FeTe$_{0.55}$Se$_{0.45}$ and Bi$_2$Te$_3$/Bi$_2$Sr$_2$CaCu$_2$O$_{8+\delta}$, non-trivial twofold pairing symmetry was observed [9,10]. However, in these two systems, the topological non-trivial superconducting characters are easily influenced by the intrinsic normal superconductivity from FeTe$_{0.55}$Se$_{0.45}$ and Bi$_2$Sr$_2$CaCu$_2$O$_{8+\delta}$, make it difficult to detect and manipulate the possible Majorana zero modes in these systems[11-14]. In this Bi$_2$Te$_3$/FeTe heterojunction, the superconductivity is limited at the interface and get rid of the influence from normal superconducting states. In this way, it is significant to study the pairing mechanism and topological properties of Bi$_2$Te$_3$/FeTe interface superconductor.

In previous studies, it was shown that the superconductivity in the Bi$_2$Te$_3$/FeTe heterostructure is fully generated when Bi$_2$Te$_3$ is just one quintuple layer (QL) thick and that the superconductivity is generated at the interface [6,15,16]; it was also shown that the interface superconductivity has clear two-dimensional superconducting characters[6]. The nonreciprocal transport results provide the evidence of the coexistence of superconductivity and spin-polarized surface states [17]. However, until now, there has been no comprehensive study on the superconducting gap symmetry of the Bi$_2$Te$_3$/FeTe system, which is crucial to the understanding of its microscopic pairing mechanism and may provide insightful clues for its potential to achieve 2D topological superconductor at this interface [18,19].

In this study, we investigate the gap symmetry in this interface superconductor by QPI and electrical transport measurement using low-temperature scanning tunneling microscopy (LT-STM) and physical property measurement system (PPMS). Our QPI results inside and outside the superconducting gap provide clear evidence of an anisotropic superconductivity gap symmetry in this system. In addition, the transport measurement is consistent with the QPI results and reveals a twofold superconducting gap symmetry.

Firstly, we demonstrate the two-dimensional superconductivity in 3 QL $Bi_2Te_3$ islands using LT-STM. Figure 2(a) shows an STM topographic image of the 3 QL $Bi_2Te_3$ islands, while a typical dI/dV spectrum taken over this island is shown in Figure 1(b), with two coherent peaks located at around ± 1.2 mV, similar to what we reported before [15]. In order to study the vortex property, we performed the dI/dV imaging with a magnetic field applied perpendicular to the interface at 1.5 mV, which corresponds to the energy of the coherence peaks in 2 QL $Bi_2Te_3$, which is typically used in vortex imaging. This is to locate the vortices' positions, since the uprise of the ZBCP in the centre of the vortex will reduce the intensity of the coherence peaks, resulting in a depression in the dI/dV imaging. The dI/dV imaging at 1.5 mV was taken on the same area as Figure 2(c) under magnetic fields of 0.5 T, the vortices can be clearly observed on both 2 QL and 3 QL $Bi_2Te_3$ area. Each of them separates from the others by about 15 to 30 nm. In many unconventional superconductors, the shape of a vortex can be closely related to the gap symmetry of a superconductor [9,20]. In our system, we also observed slightly elongated or almost round vortices on all superconducting $Bi_2Te_3$ area even at 1 QL $Bi_2Te_3$ island.(Supplementary Materials Fig.S1) But the shape of vortex is easily influenced by crystal structure, interaction and step edges [21], it is hard to give insightful clues for pairing symmetry by the shape of vortex in this $Bi_2Te_3$/FeTe interfacial superconductor due to the clear influence from surrounding environment such as terrace size or step edges.

In order to reveal the gap symmetry of $Bi_2Te_3$/FeTe interfacial superconductor, we performed the quasiparticle interference measurement outside and inside the superconducting gap. Quasiparticle interference is an effective method to investigate the pairing symmetry and has been widely used in different superconducting systems such as LiFeAs, Sn/Si(111), $AV_3Sb_5$ (A=K, Rb, Cs), $Bi_2Te_3$/Fe (Te, Se), and $Bi_2Te_3$/$Bi_2Sr_2CaCu_2O_{8+\delta}$ (Bi-2212) [9,10,22-24]. All QPI experiments were carried out in ultra-high vacuum conditions in order to have a pristine sample surface. We performed the dI/dV mapping at various energies (sample biases), outside and inside the superconducting gap. Figures 3(a), (c) and (d) are Fourier-transformed dI/dV images at biases of 10 mV, 4 mV, and 1.0 mV, respectively. At biases of 10 mV and 4 mV, scattering patterns with six spots along the $\Gamma - M$ directions (see Figure 3(a)) are clearly observed, two of which (labeled 1 and 4) having higher intensities than the other four. These scattering spots are mainly formed by the $q_2$ vector, as shown in the schematic in Supplementary Figure S2(a) [25-28]. At energies below the superconducting gap of around 1.0 meV, only spots 1 and 4 are discernable, such as Figure 3(d) at 1.0 mV. The other four scattering spots, which are observed at higher energies, is not discernible. This can also be seen in other Fourier-transformed dI/dV images at biases of 1.0 mV, 0.5 mV, 0.0 mV, −0.5 mV, −1.0 mV, and −1.5 mV, as shown in Supplementary Figure S3.

Before any conclusion is drawn, we first show that the disappearance of four spots of 2, 3, 5, and 6 is less likely due to their intensities becoming too weak (i.e., comparable to the background signal). The ratio of the maximum intensities between the three pairs of bright spots ($I_{1,4}$: $I_{3,6}$: $I_{2,5}$) are around 1:0.3965:0.2625, as shown by the line profiles in Supplementary Figure S2(c). Assuming the ratios between the three pairs of spots in Supplementary Figure S2(d) are the same as those spots in Supplementary Figure S2(c), with the maximum intensity of spots 1 and 4 in Figure S2(f) being around 10.175 as measured in Figure S2(d), the expected intensity of spots 3 and 6, and spots 2 and 5, should be around 3.93 and 2.67, respectively, which corresponds to the two horizontal lines in Figure S2(f). The intensities of spots 3 and 6 are well above the background signal and should be clearly visible, as no significant change in scattering rate that generates these scattering spots is expected. Though the intensities of spots 2 and 5 are slightly weaker, they are still above the background signal and these two spots should be visible as well. The fact that these four scattering spots are not observed below the superconducting gap strongly suggests that the original spots that correspond to these scattering spots are absent at these energies. This means that quasiparticle density of states is not present or substantially weakened at these spots at these energies, suggesting the opening of superconducting gap in the corresponding wave vectors in the k-space. Since there is no simultaneous disappearance of spots 1 and 4, the superconducting gap has not opened for the wave vectors corresponding to these two spots. This suggests that the superconducting gap function is anisotropic. QPI pattern with two spots has been observed before on $Bi_2Te_3$/$FeTe_{0.55}Se_{0.45}$ [9], and also with four spots on $Bi_2Te_3$/$Bi_2Sr_2CaCu_2O_{8+\delta}$ [10].

In order to further explore the gap symmetry of the $Bi_2Te_3$/FeTe system, we also carried macroscopic transport measurements. The temperature dependence of resistance (R-T) plotted in Figure S4(b) shows a relatively sharp transition from normal state to superconducting state, with the superconducting transition temperature ($T_c$) around 13 K. This transition temperature is consistent with previous transport and STM/STS work [6,15] and indicates the high-quality of our sample. Figure S4(c) displays the R-T curves under different in-plane magnetic fields ($H^{//}$). The previous works carried by He et al. and Yasuda et al. also confirms the superconductivity is limited to the interface and shows two-dimensional characteristics [6,17,29-32]. These characteristics make the $Bi_2Te_3$/FeTe material system particularly suitable for exploring the intrinsic interfacial superconducting property, as the two compounded materials are non-superconducting.

We performed the in-plane angle-dependent magnetoresistance (AMR) measurement at temperatures between $T_c^{onset}$ and $T_c^{zero}$ to investigate the superconducting pairing symmetry at the macroscopic scale [33-39]. As the magnetic field applied in the a-b plane of the sample (parallel to the interface) rotates around c axis (perpendicular to the interface), we observed a pronounced twofold variation of magnetoresistance from $\theta = 0°$ to $\theta = 360°$ ($\theta = 0°$ corresponds to in-plane magnetic field along the a or b axis of FeTe) in the superconducting transition region. This twofold variation is robust and well described by cosine function with the form $\cos(2\theta + \phi)$ when we varied the temperature or magnetic field, as shown in Figures 4(a-d). The angle-dependent magnetoresistances at a magnetic field of 12 T are shown in Figures 4(a) and 4(c). For temperatures measured below $T_c^{onset}$, a clear twofold variation is observed. On the contrary, no clear variation is observed at 20 K, which is above the superconducting transition temperature. This implies that the twofold symmetry in the angle-dependent resistance is a characteristic of the superconducting state, instead of the normal state. This also suggests that the observed twofold symmetry is less likely due to structural variation, unless there is a structural transition below 20 K, which has not been observed before in $Bi_2Te_3$ or FeTe [40,41]. The same twofold symmetry is observed when the magnetic field is varied from 2 T to 12 T at 12.5 K (which is within the transition region), as shown in Figures 4(b) and 4(d).

In the angle-dependent magnetoresistance measurement, it is often difficult to ensure a perfect alignment between the magnetic field and the rotating sample a-b plane. However, as shown by Hamill et al. [37]. and Xiang Ying et al. [39], the feature of the twofold symmetry is essentially not affected by the misalignment. Moreover, the current induced movement of vortices may also lead to the observation of a twofold symmetry in the magnetoresistance in superconducting states but can be ruled out by using different current directions in the measurement. These two possible extrinsic effects are also discussed in Supplementary Materials in detail.

In order to have a deeper understanding of the twofold symmetry of the measured angle-dependent magnetoresistance, we carried out a series of measurement of angle-dependent R-T under different magnetic-field strengths and extracted the zero-temperature upper critical field ($H_{C2}$) following the same formula in a previous work [6]. The results are shown in Figure 4(e), which clearly shows twofold symmetry, similar to that of the angle-dependent magnetoresistance. The variation of $H_{C2}$ has a $\pi/2$ phase shift compared with magnetoresistance, i.e., a larger $H_{C2}$ corresponds to smaller magnetoresistance at the same angle. This is expected, as a larger resistance needs a larger magnetic field to suppress.

Overall, the anisotropic magnetoresistance and $H_{C2}$ behavior indicate the anisotropic coherent length inside the superconducting states in $Bi_2Te_3$/FeTe interfacial superconductor and indicate an anisotropic superconducting gap function.

Next, we will briefly discuss the possible origin of the observed twofold gap symmetry. Although both the QPI and transport results indicate a possible twofold gap symmetry, it may simply be induced by the lattice mismatch. This C2 symmetry relies on the crystal orientation of $Bi_2Te_3$ and can be modulated by the two as-grown rotation domains of $Bi_2Te_3$ as shown in Figures 1(b) and (c) [16]. The lattice mismatch in $Bi_2Te_3$/FeTe heterojunction will naturally introduce an additional symmetry to the system and lead to a possible C2 symmetry as our QPI pattern shows anisotropy above the superconducting gap. This means that normal-state band is anisotropic, thereby anisotropic features should also emerge in the superconducting state. In this case, this anisotropic modulation can be detected in normal states by electronic transport measurement as the normal-state band is anisotropic, which is clearly absent in our AMR results. By the way, in another previous work, the superconductivity in monolayer FeTe grown on $Bi_2Te_3$ shows no modulation by the 8 ×2 moiré pattern induced by the lattice mismatch [42]. Both the absence of this twofold symmetry in normal states and non-modulated superconductivity in monolayer FeTe oppose the lattice mismatch effect. By this roughly discussion, it is more likely this twofold gap function is an intrinsic property of this interfacial superconductor in superconducting states.

This deduction will lead to correlation between the frustrated antiferromagnetic order (AFM) and superconductivity [43-45]. But interestingly, in our experiment, the twofold modulation in AMR in normal states is clearly absent and no coexistence of superconductivity and AFM order is observed in our previous work [42]. Besides, if this twofold modulation is strongly correlated with the bi-collinear AFM order, the twofold modulation should align with the Fe-Fe direction not the Te-Te direction, which is shown in our QPI and AMR results. It indicates that the twofold superconductivity pairing symmetry is not closely related to the AFM order , which is the main difference between our results and results in Ref.[43].

Above all, the most possible mechanisms are the interplay between superconductivity and topological surface states in this interface as the $Bi_2Te_3$ carries the spin-momentum locked surfaces states and pervious theoretic work predicts the pure FeTe may carry the topologically nontrivial states, giving the possibility to achieve homogenous Majorana zero modes in superconducting FeTe compared with $FeTe_{1-x}Se_x$ systems [46-51]. Further experiments, using lower temperature and spin-polarized STM, is clearly necessary to distinguish the discrete in-gap states and shed light on the origin of this twofold pairing symmetry, which may point out whether it related to electron-electron correlation or the spin-momentum locked superconducting states [52].

**Conclusion:**

In conclusion, we studied the superconductivity pairing symmetry in an iron-based interfacial superconductor by using both macroscopic electronic transport measurements and microscopic scanning tunneling measurements. It is clearly found from our study that the superconducting gap symmetry of $Bi_2Te_3$/FeTe shows twofold anisotropic symmetry. Our results also highlight the proposal of achieving artificial topological superconductors in Bi-based topological insulators/FeTe systems and introduce a viable platform to explore the relationship between strongly electron-correlation states and superconductivity in iron-based superconductors.

ASSOCIATED CONTENT

**Supporting Information**

Experimental Methods. STM topographic image and corresponding dI/dV images with magnetic field acquired on a 1 QL $Bi_2Te_3$ area (Figure S1). QPI pattern at different energies and intensity analysis (Figure S2). Fourier-transformed dI/dV images at biases below the superconducting gap (Figure S3). Electronic transport measurement of $Bi_2Te_3$/FeTe heterojunction (Figure S4). Analysis of Extrinsic Effects. The R-T curve and angle-dependent magnetoresistance curve of sample No.2 (Figure S5)

# AUTHOR INFORMATION

**Acknowledgements**: We thank Hong Ding and Zhuoyu Chen for the fruitful discussions. This work was supported by the National Key Research and Development Program of China (No. 2018YFA0307100), the National Natural Science Foundation of China (Nos. 61734008 and 11774143), Guangdong Provincial Key Laboratory (Grant No.2019B121203002,), Technology and Innovation Commission of Shenzhen Municipality (JCYJ20170412152334605, KQJSCX20170727090712763 and ZDSYS20190902092905285). J.W.M was partially supported by the program for Guangdong Introducing Innovative and Entrepreneurial Teams (No. 2017ZT07C062). The transport data were obtained using equipment maintained by Southern University of Science and Technology Core Research Facilities.

***Emails of corresponding authors**: meijw@sustech.edu.cn, wangg@sustech.edu.cn.

‡**These authors contributed equally to this work**.

## Competing interests

The authors declare no competing financial interest.

## ABBREVIATIONS

ZBCP, zero bias conductance peak; QL, quintuple layer; LT-STM, low-temperature scanning tunneling microscopy; AFM, antiferromagnetism; RHEED, Reflection High Energy Electron Diffraction.

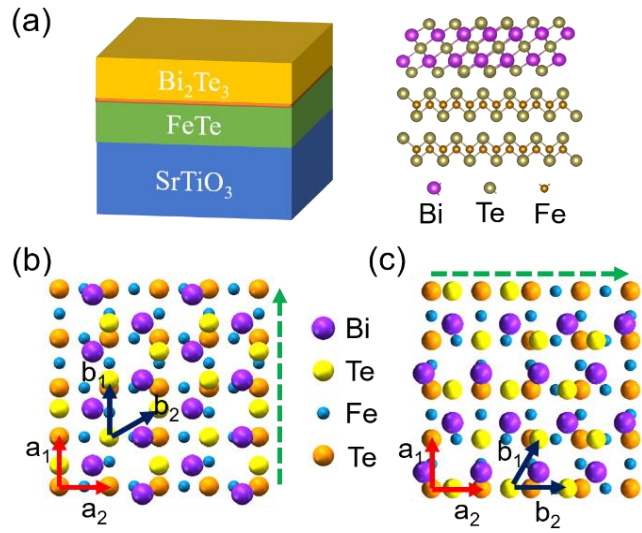

**Figure 1** Structure of $Bi_2Te_3$/FeTe heterojunction. (a) Side view of schematic picture of $Bi_2Te_3$/FeTe heterojunction with atomic structure. (b) and (c) Schematic image of two different as-grown domains of $Bi_2Te_3$ with base vector ($a_1$, $a_2$ for FeTe and $b_1$, $b_2$ for $Bi_2Te_3$), the green dash line shows the alignment direction between $Bi_2Te_3$ and FeTe.

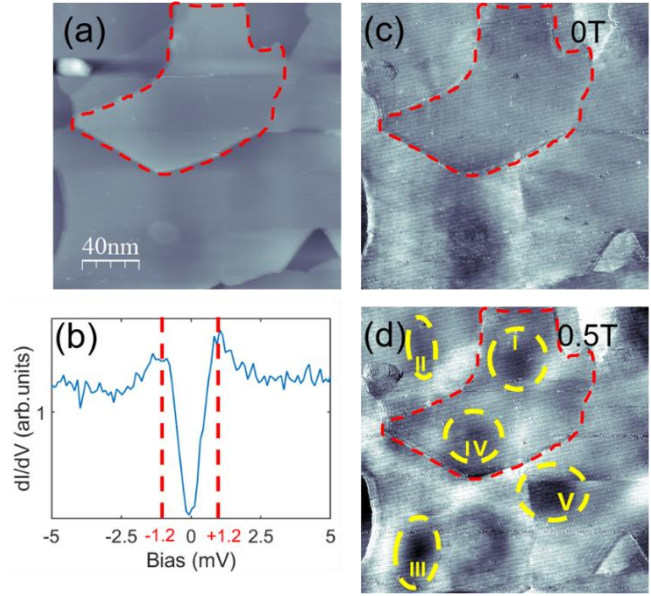

**Figure 2 STM/STS in 3 QL Bi$_2$Te$_3$** (a) STM topographic image acquired on a 2 or 3 QL Bi$_2$Te$_3$ area at 1.1 K. ($V_{Bias}$ = 1.8 mV, $I_{Tunnel}$ = 30 pA) Dashed line outlines the 3 QL Bi$_2$Te$_3$ island. (b) Typical dI/dV spectrum acquired on 3 QL Bi$_2$Te$_3$ island. Red dashed lines mark the position of coherent peak. (c) STS imaging of (a) at 0T. ($V_{Bias}$ = 1.5mV, $I_{tunnel}$ = 5pA ) (d) Vortices mapping of (a) at 0.5T. Yellow dashed line outlines different vortices (Roman number 1~5, setpoint: $V_{Bias}$ = 1.5 mV, $I_{Tunnel}$ = 5 pA). Note the vortices appear dark in this dI/dV image because it was acquired at a bias around the superconducting gap magnitude, instead of zero-bias.

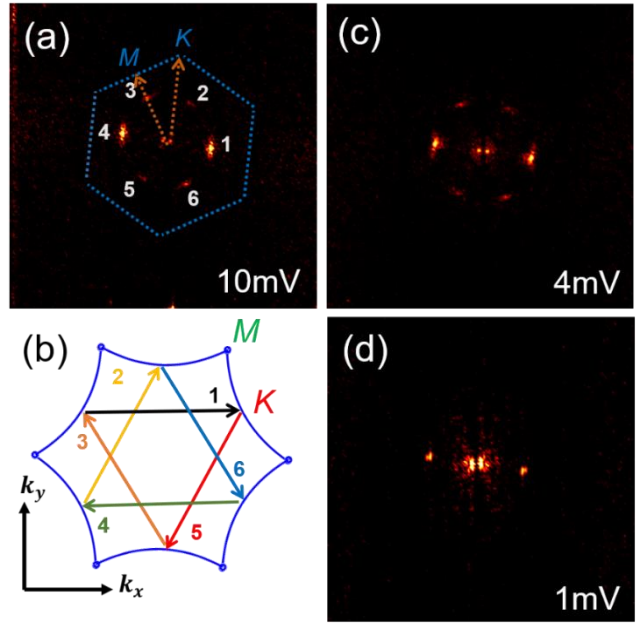

**Figure 3** Anisotropic superconducting gap resolved by QPI pattern. (a) Fourier-transformed dI/dV image at 10mV with the 2D schematic image of the first Brillouin zone (blue dashed area). (b) 2D schematic image of hexagonal Fermi surface and possible $q_2$ scattering vectors. (c-d) Fourier-transformed dI/dV images at 4mV (outside SC gap) and 1.0mV (inside SC gap). Spots 2,3,5,6 are absent inside SC gap.

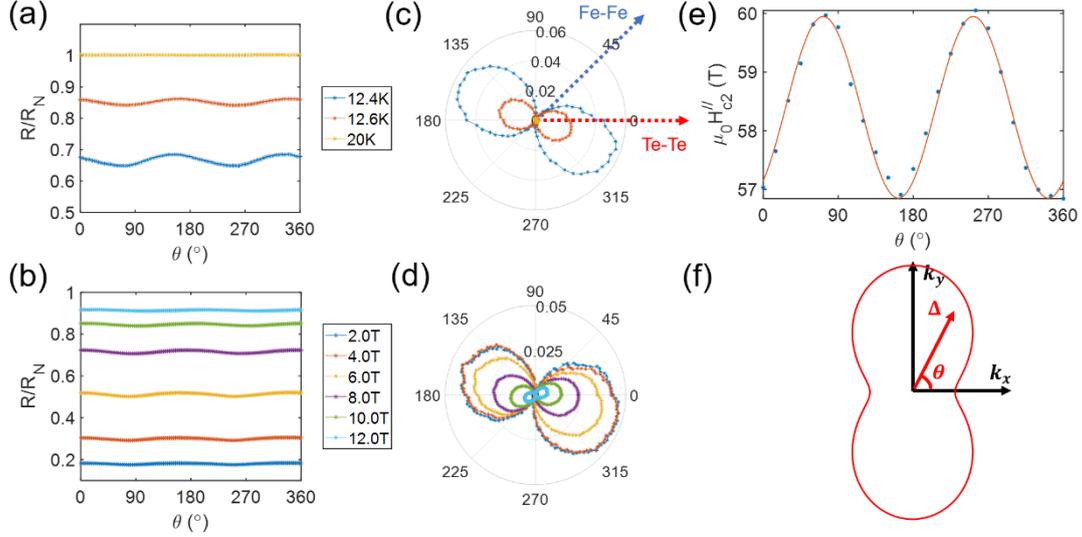

**Figure 4 Twofold modulation resolved by angular magnetoresistance measurement (AMR).** (a) AMR at 12 T measured at different temperatures indicated in the figure, normalized by $R_N$ (5.05 Ω, resistance at 20 K, 0 T). (b) AMR at 12.5 K under different magnetic fields, normalized by $R_N$ (5.05 Ω, resistance at 20 K, 0 T). (c) Polar plot of the data ratio in (a), normalized by resistance at normal states (20 K), $R_{normalized} = (R-R_{min})/R_N$. (d) Polar plot of the data ratio in (b), normalized by resistance at normal states (20 K), $R_{normalized} = (R-R_{min})/R_N$. (e) Angular dependence of in-plane upper critical field, the $H_{c2}$ data is fitted by Landau-Ginzburg formula. (f) Schematic image of resultant angular dependence of anisotropic SC gap function in k-space.

**Supporting Information**

Supporting Information of "Twofold Symmetry Observed in Bi$_2$Te$_3$/FeTe Interfacial Superconductor"

Xinru Han[1, ‡], Hailang Qin[1, ‡], Tianluo Pan[1, 3, ‡], Bin Guo[1], Kaige Shi[1], Zijin Huang[1], Jie Jiang[1], Hangyu Yin[1], Hongtao He[1,2], Fei Ye[1,3], Wei-Qiang Chen[1,3], Jie-Wei Mei[1, 3, *], Gan WANG[1, 3, *]

[1]Department of Physics, Southern University of Science and Technology, Shenzhen 518055, China

[2]Shenzhen Key Laboratory for Advanced Quantum Functional Materials and Devices, Southern University of Science and Technology, Shenzhen 518055, China

[3]Guangdong Provincial Key Laboratory of Quantum Science and Engineering, Shenzhen Institute for Quantum Science and Engineering, Southern University of Science and Technology, Shenzhen 518055, China

**\*Emails of corresponding authors:** meijw@sustech.edu.cn, wangg@sustech.edu.cn.

‡These authors contributed equally to this work.

**This file includes:**
- **Section 1: Experimental Details**
- **Fig. S1:** STM topographic image and corresponding dI/dV images with magnetic field acquired on a 1 QL Bi$_2$Te$_3$ area.
- **Fig. S2:** QPI pattern at different energies and intensity analysis.
- **Fig. S3:** Fourier-transformed dI/dV images at different biases.
- **Fig. S4:** Electronic transport measurement of Bi$_2$Te$_3$/FeTe heterojunction.
- **Section 2: Analysis of Extrinsic Effects**
- **Fig. S5:** The R-T curve and angle-dependent magnetoresistance curve of sample No. 2.

**Section 1: Experimental Details**

The substrate is a single-crystal 0.7%-wt Nb-doped SrTiO$_3$(001) (STO) (Hefei Kejing). The FeTe thin films were grown by co-evaporation of Fe (99.995%) and Te (99.999%) sources onto the substrate held at around 300 ˚C, followed by the deposition of Bi$_2$Te$_3$ thin films, by evaporating Bi$_2$Te$_3$ compound source with the substrate at about 225 ˚C. The STM measurements were performed at 1.1 K (unless otherwise specified) with etched tungsten tips with the bias voltage applied to the sample. A magnetic field perpendicular to the sample interface is applied during vortex imaging.

The grown samples were then transferred *in-situ* to the SPECS Joule-Thomson (JT) LT-STM system. The base pressure in the MBE chamber and STM chamber was both about $2 \times 10^{-10}$ mbar, respectively. The STM tips were sputtered with an $Ar^+$ ion sputter gun and tested on a reference Au(111) single-crystal. Topographic images were acquired in constant-current mode and dI/dV spectra were measured by using a standard lock-in method with a modulation frequency of 998 Hz and a typical modulation amplitude of 0.1 – 0.5 mV depending on the bias voltage range of the spectra. Each spectrum is usually an average of tens of or more spectra acquired at the same location. The STM images were analyzed with WSXM software[1]. Brighter in color in an STM image means higher in value in this study. The thickness of the FeTe film is about 40 nm.

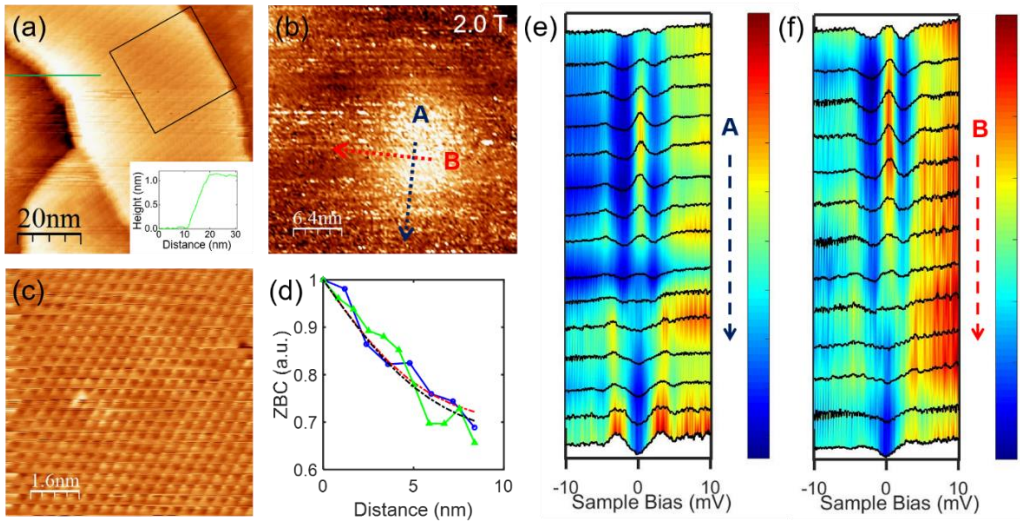

**Figure S1 STM topographic image and corresponding dI/dV images with magnetic field acquired on a 1 QL Bi2Te3 area.** (a) STM image of a 1 QL $Bi_2Te_3$/FeTe area at 1.1 K and 2.0 T. ($V_{Bias}$ = 2.5 mV, $I_{Tunnel}$ = 50 pA) (b) Zero bias dI/dV image of an area within (a). (setpoint: $V_{Bias}$ = 5 mV, $I_{Tunnel}$ = 50 pA) (c) Typical atomic resolution image taken in the place where the vortex is held. ($V_{Bias}$ = 6 mV, $I_{Tunnel}$ = 600 pA) (d) Normalized ZBC profiles taken from the lower part of the spectra in (e) (circle) and (f) (triangle), respectively. The dashed lines are the corresponding fittings. (e) A series of equally-spaced dI/dV spectra acquired along the blue dashed line A, as shown in (b). ($V_{Bias}$ = 10 mV, $I_{Tunnel}$ = 50 pA (f) A series of equally-spaced dI/dV spectra acquired along the red dashed line B as shown in (b). ($V_{Bias}$ = 10 mV, $I_{Tunnel}$ = 50 pA)

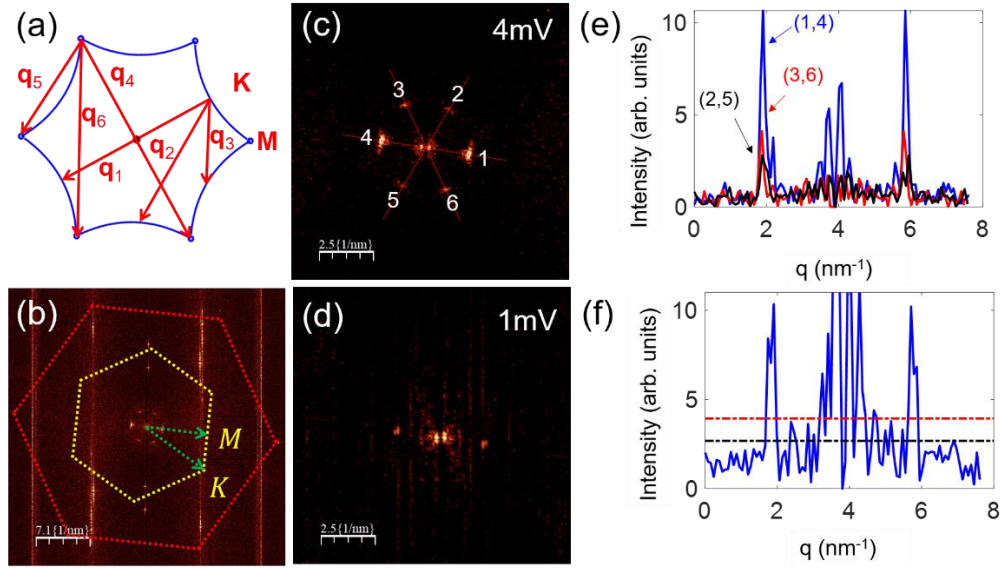

**Figure S2 QPI pattern at different energies and intensity analysis.** (a) A schematic showing the possible scattering vectors within a constant-energy contour (CEC) of $Bi_2Te_3$ (b) Full image of the Fourier-transformed dI/dV image at 4 mV. The red hexagon indicates the Bragg spots from the top Te atoms on $Bi_2Te_3$, and the yellow hexagon shows the first Brillouin zone. (c)-(d) Zoom-in view of the Fourier-transformed dI/dV images at various biases of 4 mV, and 1.0 mV, respectively. (e). (f) Line profile across spots 1, 4, spots 3, 6, and spots 2, 5 in (c). (f) Line profiles across spots 1 and 4 in (d). The upper and low dashed lines corresponding to the expected intensities of spots 3, 6, and spots 2, 5, if assuming the same intensity ratios among the six spots as those in (b).

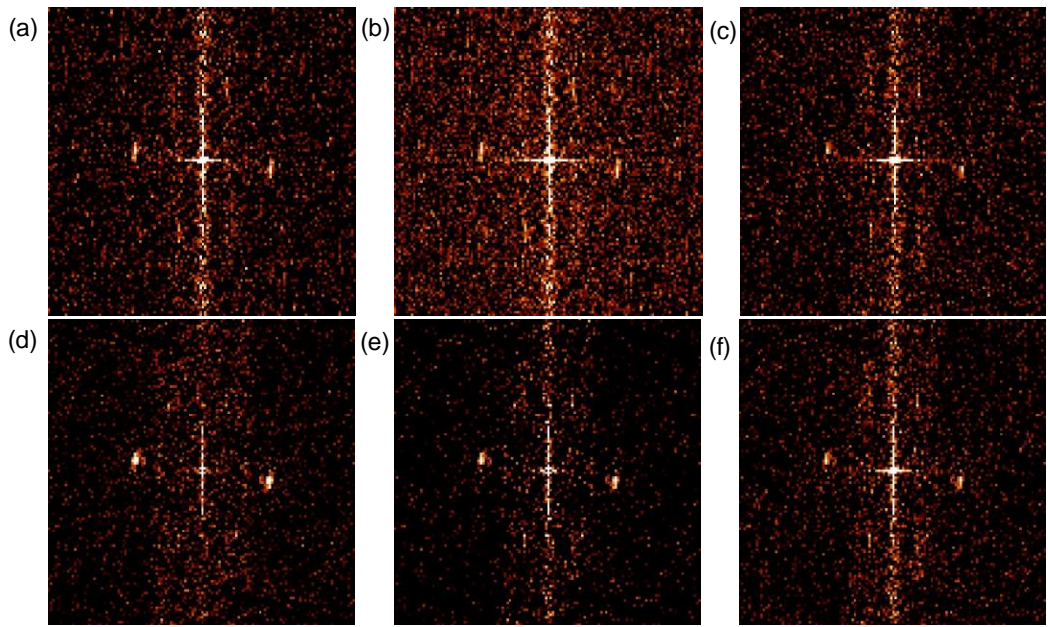

**Figure S3 Fourier-transformed dI/dV images.** (a-f) Fourier-transformed dI/dV images at biases of 1.0 mV, 0.5 mV, 0.0 mV, −1.5 mV, −1.0 mV, and −0.5 mV, respectively.

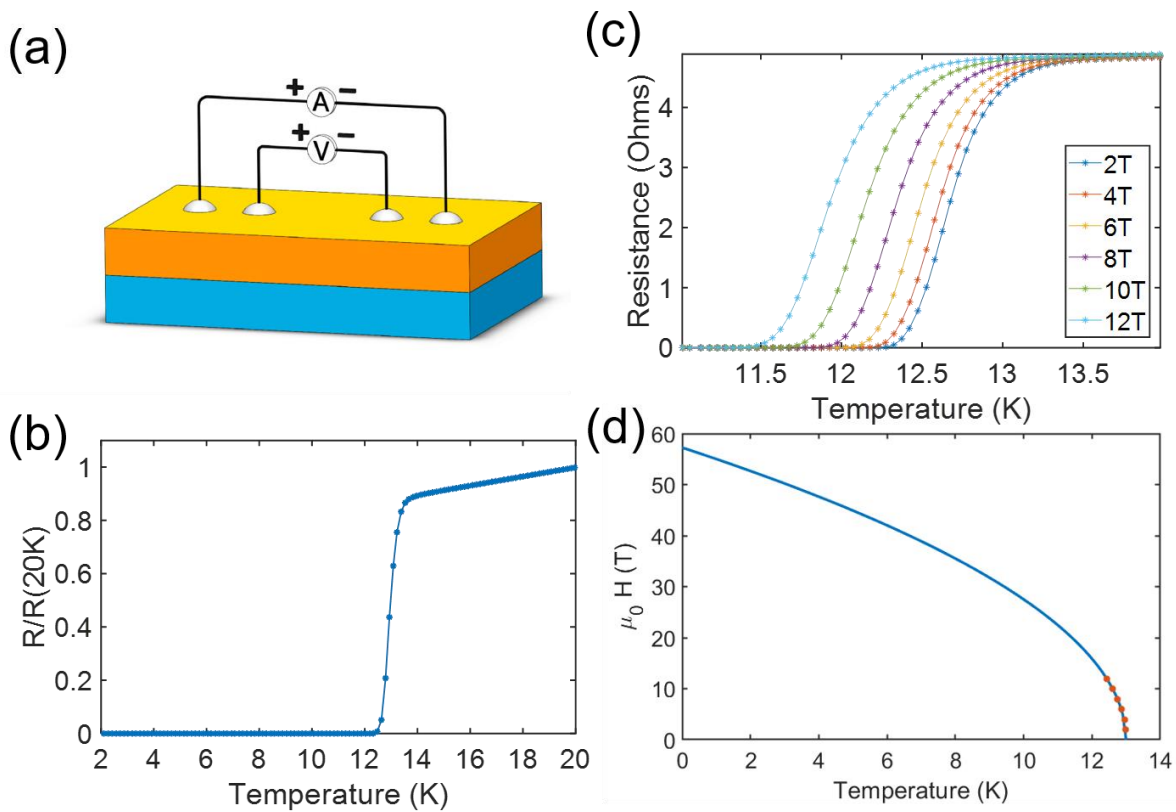

**Figure S4 Electronic transport measurement of Bi$_2$Te$_3$/FeTe heterojunction** (a) The schematic of the Bi$_2$Te$_3$/FeTe device in our measurement set-up. (b) R-T curve of the device with zero magnetic field. (c) R-T curves of Bi$_2$Te$_3$/FeTe under different in-plane magnetic fields (2 T, 4 T, 6 T, 8 T, 10 T, 12 T) applied in the same direction. (d) Fitting in-plane H$_{c2}$ according the data shown in (c) by Landau-Ginzburg formula.

**Section 2: Analysis of Extrinsic Effects**

In this section, we will discuss two kinds of extrinsic effects that can lead to a two-fold modulation in this experiment.

**1. The out-of-plane magnetic field caused by tilt.**

We discuss the possibility of the effect induced by the out-of-plane component of the applied field, which can cause two-fold modulation in magnetoresistance due to an unintentional cant of the sample plane with respect to the sample rotation plane. Considering this canting effect is hard to eliminate, we simply simulate the condition as the superconducting symmetry couples with this canting effect and observed that the varied strength of canting effect can lead to a phase shifting (Figures not shown) which is also observed in other system such as few-layer NbSe$_2$[2]. This phase shifting is consistent with our resistance data shown in the main text and confirms that a two-fold superconducting symmetry exist in the Bi$_2$Te$_3$/FeTe system as we can change the strength of

applied magnetic field to change the relative strength between the superconductivity and out-of-plane effect

2. **The current induced vortices moving.**

Another possible cause is the vortices moving[3]. In order to rule out this extrinsic effect, we measured another sample using the same method. For sample No. 2, we performed transport with the same current direction as sample No. 1 shown in main text but get a different modulation alignment as the minimum value located at zero position, having a $\pi/2$ shifting compared with sample 1, as the twofold have the possibility to align with a or b axis of FeTe. This different alignment also shows this twofold modulation is not induced by the current.

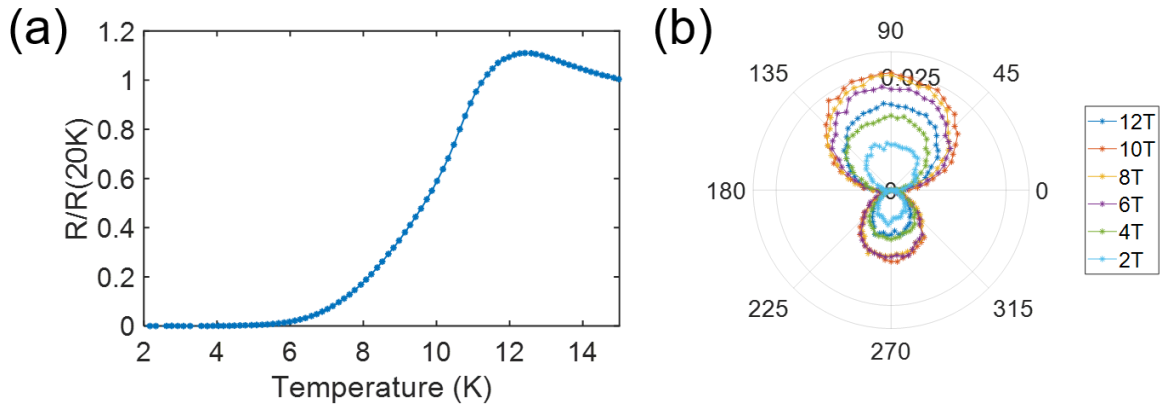

**Figure S5 The R-T curve and angle-dependent magnetoresistance curves of sample No. 2** (a) The normalized resistance vs. temperature curve of sample No. 2. (b) The angle-dependent magnetoresistance curves of sample No. 2 with different magnetic fields (2 T, 4 T, 6 T, 8 T, 10 T, 12 T).